\documentclass[preprint, prb , showpacs]{revtex4}

\usepackage{graphicx,epsfig, setspace}
\usepackage{dcolumn}
\usepackage{amsmath}

\newcommand{\vect}[1]{\ensuremath{\boldsymbol{#1}}}

\begin{document}

\preprint{}
\title{Reducing the critical switching current in nanoscale spin valves}
\author{Jan \surname Manschot}
\affiliation{Kavli Insitute of Nanoscience, Delft University of Technology, 2628 CJ
Delft, The Netherlands and Department of Physics, Norwegian University of Science and Technology,
N-7491 Trondheim, Norway}
\author{Arne \surname Brataas}
\affiliation{Department of Physics, Norwegian University of Science and Technology,
N-7491 Trondheim, Norway}
\author{Gerrit E. W. \surname Bauer}
\affiliation{Kavli Insitute of Nanoscience, Delft University of Technology, 2628 CJ
Delft, The Netherlands}

\begin{abstract}
The current induced magnetization reversal in nanoscale spin valves is a
potential alternative to magnetic field switching in magnetic memories. We show that the critical switching current can be decreased
by an order of magnitude by strategically distributing the resistances
in the magnetically active region of the spin valve. In addition, we
simulate full switching curves and predict a new precessional state. 
\end{abstract}

\pacs{73.63.-b, 75.47.-m, 75.70.Ak, 85.75.-d}
\maketitle

The prediction that a spin-polarized current can excite and reverse a
magnetization\cite{slonczewski,berger} has been amply confirmed by recent
experiments.\cite{kent, kiselev} The current-induced magnetization
dynamics is interesting as an efficient mechanism to write information
into magnetic random access memories as well as to generate
microwaves.\cite{rippard} Unfortunately, the critical currents for
magnetization reversal are still unattractively high.\cite{sun4} In this Letter, we apply a previously developed
microscopic formalism\cite{brataas2} to understand the critical
current in spin valves quantitatively and propose a strategy to reduce it by up to
an order of magnitude. We also solve the micromagnetic equations with
accurate angle-dependent magnetization torque and spin-pumping\cite%
{tserkovnyak2} terms and predict switching to a novel precessional state.

We will first consider a generic F(erromagnetic)$|$N(ormal)$|$F
spin valve biased by a voltage difference $V$. The two ferromagnetic
reservoirs are assumed to be monodomain; the magnetizations differ by
an angle $\theta$. Charge and spin currents excited by an applied bias
can be calculated accurately by magneto-electronic circuit
theory\cite{brataas2} with parameters determined from first-principle
calculations\cite{xia} that agree well with experimental
data.\cite{bauer} To this end we 
dissect the pillar into three nodes (the reservoirs and the normal
metal) connected by two, not necessarily identical resistive elements $G_{\mathrm{L}}$ and
$G_{\mathrm{R}}$. Each of them is
characterized by the conductances $g=g^{\uparrow\uparrow}+g^{\downarrow\downarrow}$, the polarization
$p=(g^{\uparrow \uparrow }-g^{\downarrow \downarrow })/g$ and the normalized mixing
conductance $\eta =2g^{\uparrow \downarrow }/g$. $g^{\uparrow \uparrow}$ and $g^{\downarrow \downarrow }$ are, respectively, the conductances for
electrons with spin parallel and anti-parallel to the magnetization and $%
g^{\uparrow \downarrow }$ is the material parameter that governs the
magnetization torque. The magnetically active region includes layers of
thickness up to the spin-flip diffusion lengths from the interfaces. Any
resistance outside this region is parasitic and not considered here. The
conductances are effective parameters determined by the resistance of the
ferromagnetic and normal metal bulk, that of the interfaces to the normal metal and the
resistance of an eventual outer normal metal that fits into the magnetically
active region. For simplicity we disregard the bulk resistance of the normal
metal island and the imaginary part of $g^{\uparrow \downarrow }$ (for
metallic interfaces smaller than $10\,\%$ \ of the real part\cite{xia,
stiles}).

The transverse component of the spin current is absorbed in the
ferromagnet \cite{stiles} and the associated spin-transfer torque can
excite the magnetization.\cite{slonczewski,berger} Circuit theory has
been used to derive analytic expressions for the torques in a
symmetrical spin valve as function of the angle $\theta$ between the
magnetization directions.\cite{bauer} However, the spin-transfer
torque depends strongly on the resistance distribution, even its sign
may change with asymmetry.\cite{manschot,kovalev} In this Letter we
propose to engineer the spin valve resistance distribution in order to
minimize the critical current for magnetization reversal from the
parallel to the antiparallel configuration (the opposite process is
much less sensitive to this asymmetry). Expanding
the spin-transfer torque from circuit theory to 
first order in $\theta$, the
normalized torque $i_{\mathrm{s}}=|\boldsymbol{m}\times
(\boldsymbol{I}_{\mathrm{s}}\times \boldsymbol{m})|/I_{\mathrm{c}}$ on
the \emph{left} magnetization reads 
\begin{align}
\left. i_{\mathrm{s}}(\theta )\right\vert _{\theta \approx 0}& =\frac{\hbar 
}{2e}\left( \frac{g_{\mathrm{L}}\eta _{\mathrm{L}}}{g_{\mathrm{L}}\eta _{%
\mathrm{L}}+g_{\mathrm{R}}\eta _{\mathrm{R}}}\right) \times   
\label{eq:torappr} \\
& \frac{g_{\mathrm{R}}\eta _{\mathrm{R}}(p_{\mathrm{R}}-p_{\mathrm{L}})+g_{%
\mathrm{L}}p_{\mathrm{R}}(1-p_{\mathrm{L}}^{2})+g_{\mathrm{R}}p_{\mathrm{L}%
}(1-p_{\mathrm{R}}^{2})}{g_{\mathrm{L}}(1-p_{\mathrm{L}}^{2})+g_{\mathrm{R}%
}(1-p_{\mathrm{R}}^{2})}\theta.   \notag
\end{align}
In an asymmetric structure, $p_{\mathrm{R}}-p_{\mathrm{L}}\neq 0$, the slope
of the spin torque can be considerably enhanced compared to the torque
in a symmetric structure $i_{\mathrm{s}}=(\hbar /2e)(p/2)\theta$.

We investigate a realistic (but non-unique) model for the asymmetry by
extending the layer sequence from F$|$N$|$F to N$_{1}|$F$|$N$|$F$|$N$_{2}$.
We take here the four N$|$F interfaces to be equal and assume that the
ferromagnets are thin enough that the bulk contribution is negligibly small.
Numerical results require values for the interface resistances that have
been measured accurately in the current-perpendicular-to-plane 
geometry.\cite{pratt, gijs} We adopt here Co/Cu interfaces with
cross-section $1.26 \times 10^{4}\,\mathrm{nm^{2}}$, whence $1/G=0.0183 \, \Omega $, 
$p=0.75$ and $\eta =0.38$.\cite{yang,xia} The asymmetry is modeled by the
normal metal sandwich \emph{outside} the symmetric F$|$N$|$F structure.
The conductivity of the resistive element connecting the left
(right) reservoir to the adjacent normal metal layer is $G_{1}$ ($G_{2}$).
The asymmetry is expressed by varying the values for $G_{1}$ and $G_{2}$ for
constant series resistance $1/G_{1}+1/G_{2}=0.37\,\Omega$. 
The total collinear pillar resistances can now be calculated by the two
current model to be 0.505 and $0.534\,\Omega$, which are typical
values for recently fabricated nanopillars.\cite{kent} We assume 
that the right magnetic layer is magnetically hard and is treated as static
\textquotedblleft polarizer\textquotedblright .

The effective conductance parameters, ($\hat{g}_{\mathrm{L}}$%
, $\hat{p}_{\mathrm{L}}$ and $\hat{\eta}_{\mathrm{L}}$) for the left hand side of the pillar consisting of the
ferromagnet and the outer normal metal can be calculated in terms of the
normal metal conductance $g_{1}$ as
\begin{align}
 & \hat{g}_{\mathrm{L}}=\frac{1}{2}\left(\frac{g_{1}g^{\uparrow
      \uparrow}}{g_{1}+g^{\uparrow \uparrow}}+\frac{g_{1}g^{\downarrow
     \downarrow}}{g_{1}+g^{\downarrow \downarrow}} \right),  \qquad  \hat{\eta}_{\mathrm{L}}=\frac{2g^{\uparrow \downarrow }}{\hat{g}_{\mathrm{L}}},  
\label{eq:effparameter}  \\
 & \hat{p}_{\mathrm{L}}=\frac{g_{1}^{2}(g^{\uparrow \uparrow }-g^{\downarrow
\downarrow })}{2(g_{1}+g^{\uparrow \uparrow })(g_{1}+g^{\downarrow
\downarrow })\hat g_{\mathrm{L}}}. \notag
\end{align}
$g_{1}$ should be replaced by $g_{2}$ to obtain expressions for the right
hand side. We parametrize Eq. (\ref{eq:torappr}) as $i_{\mathrm{s}}=(\hbar
/2e)k\theta$, where numerical results for the torque
parameter $k$ are given in Table \ref{tab:k} for different distributions of
the resistance over $G_{1}$ and $G_{2}$. The dependence of the 
torque on $G_{1}:G_{2}$ can be understood simply in terms of the spin
accumulation in the parallel configuration. In contrast to a symmetric
structure, it does not vanish in asymmetric valves and can have either
sign.\cite{manschot} The additional spin accumulation that is excited
when $\theta$ becomes finite increases an antiparallel accumulation,
and thus the torque, when the extra resistance is on the left side. In
the opposite case, the initially positive spin accumulation is
cancelled at a certain angle at which the angular magnetoresistance is
minimal and the torque vanishes. \cite{manschot}  

For our specific example the spin-transfer torque on the left
magnetization is enhanced by a factor of five when all normal
resistance resides on the side of the \emph{left} magnetic
layer. The torque can thus be maximized by placing a
material with a small spin flip length (\textit{e.g.} platinum)
adjacent to the \emph{right} magnetic layer, as well as a material with  a
large spin flip length (\textit{e.g.} copper) to the left layer. The
magnitude of the torque is enhanced as well when all resistance is
placed on the other side, but its sign is changed. As shown below, in
this configuration a reversed current induces switching to a finite angle.

The magnetization dynamics is described by an extended form of the
Landau-Lifshitz-Gilbert equation.\cite{landau, gilbert} As in previous
simulations\cite{sun2,li} we adopt a single-domain model, but
we take into account accurate angle-dependent magnetization torques,\cite%
{manschot} as well as the \textquotedblleft dynamic
stiffness\textquotedblright \cite{tserkovnyak2} due to spin pumping.\cite%
{tserkovnyak3} We take the layers to be in the $y-z$ plane and the $x$-axis
in the current direction. A uniaxial effective field,
$\boldsymbol{B}_{\mathrm{eff}}$, and the fixed magnetization are
chosen parallel to the $z$-axis. Disregarding dipole and exchange
coupling between the magnetic layers, both magnetizations in the
ground state point along the external field. Analytic estimates of the
critical current are obtained here by focusing on the instability
point, at which the current-induced torque exactly equals the 
damping torque $D(\theta )$. In the presence of in-plane fields,
the critical current for complete switching does not necessarily agree
with the instability  point.\cite{valet} We disregard this
complication as well as temperature induced fluctuations of the
magnetizations since they do not interfere with the
effect of the distributed resistance.  

The maximal viscous damping reads to lowest order in small angles in
$\theta$ from the parallel configuration:  
\begin{equation}
\left. D(\theta )\right\vert _{\theta \approx 0}=\alpha M_{1} \vert\vect{B}_{\mathrm{eff}}\vert\theta,
\end{equation}
\noindent where $\alpha $ is the Gilbert damping parameter and $M_{1}$ is the magnitude of the left magnetic moment. We obtain numerical results for $M_{1}$ with the
sample cross-section defined above, a thickness of $3\,\mathrm{nm}$ and a
saturation magnetization $M_{\mathrm{s}}=1.19\cdot 10^{6}\,\mathrm{A\,m^{-1}}
$. The critical current $I_{\mathrm{c,c}}$ is then
given by  
\begin{equation}
I_{\mathrm{c,c}}=\left. \frac{D(\theta )}{i_{\mathrm{s}}(\theta )}%
\right\vert _{\theta \approx
  0}=\frac{2e\vert\vect{B}_{\mathrm{eff}}\vert M_{1}}{\hbar }\frac{%
\alpha }{k}.
\end{equation}
The total Gilbert damping parameter $\alpha$ consists of $\alpha_{0}=0.006$%
, the bulk Gilbert damping parameter and $\Delta \alpha (\theta )$
originating from the dynamic stiffness.\cite{tserkovnyak2,tserkovnyak3} We find
for $\left. \Delta \alpha\right\vert _{\theta \approx 0}$ in the limit that the spin currents are
efficiently dissipated
\begin{equation}
\left. \Delta \alpha \right\vert _{\theta \approx 0}=\frac{\gamma \hbar}{8\pi M_{1}}\left(\frac{%
2g_{1}g^{\uparrow \downarrow }}{g_{1}+2g^{\uparrow \downarrow }}+g^{\uparrow
\downarrow }\right).
\end{equation}%
The first term in parentheses is the conductance for a transverse spin
current from the (left) ferromagnet to the left reservoir. The transverse
spins escaping to the right are dissipated in the ferromagnet when 
$\theta\approx0$, the conductance for these spins is thus $\eta g/2$.
The distribution of the resistance over $G_{1}$ and $G_{2}$ is thus of
importance as well for the magnitude of the Gilbert damping. Decreasing $%
g_{1}$ decreases the damping parameter and thus the
critical current. In Table \ref{tab:k}, the excess damping $\Delta \alpha $
is given for several resistance distributions.

The critical currents $I_{\mathrm{c,c}}$ can now be calculated assuming an
effective field of $\vert\vect{B}_{\mathrm{eff}}\vert=0.2\,\mathrm{T}$. We observe that moving
the resistance to the side of the switching layer decreases the critical
current in two ways, by decreasing the excess damping and increasing the
torque. For our specific model structure the
critical current is more than five times smaller in the most asymmetric compared to the
symmetric pillar. When all resistance resides on the right hand side of the
pillar, the lowest critical current is achieved by an opposite bias. Not
only the torque, as shown above, but also the damping is then increased.
Measured critical currents can be modeled generally well with our model
(within 10 \%) when anisotropy fields are included and, in some samples, the
dynamics of the polarizer. 

A Pt layer insertion (with very short spin-flip diffusion length)
close to the switching layer as fabricated by Kiselev \textit{et
  al.},\cite{kiselev} reduces the magnetically active resistance on
the left side. However, this paper does not report an inverse
switching as predicted here. In fact, our calculated critical currents
agree best with the experimental ones for a symmetric structure
without any resistance outside the magnetic layers of equal
thickness. This ambiguity might be caused by the limitation of our
one-dimensional model to accurately describe the three-dimensional
magnetic polarizing contact in Kiselev \textit{et al.}'s device. 

\begin{figure}[thp]
\begin{center}
\includegraphics[width=8.6cm]{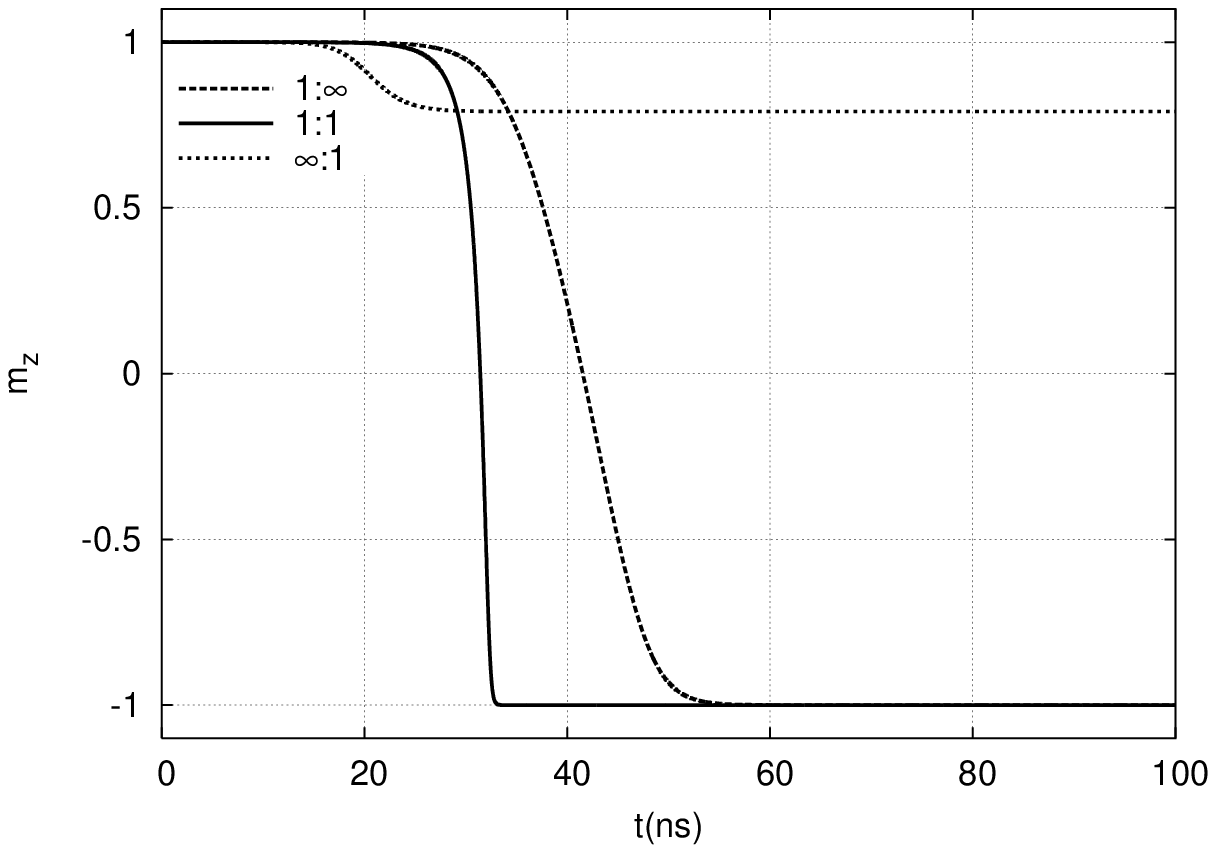}
\end{center}
\caption{Manschot}
\label{fig:asym}
\end{figure}

Finally, we present numerical simulations of complete switching curves. A
small initial torque is created by starting at $\theta _{0}=0.001$. In
Figure \ref{fig:asym} we present the switching curves $m_{z}(t)$ of the left
magnetization for three resistance distributions. $m_{z}(t)$ is the $z$%
-component of the unit vector $\boldsymbol{M}_{1}/|\boldsymbol{M}_{1}|$. All
curves are calculated with $c=I_{\mathrm{c}}/I_{\mathrm{c,c}}=1.5$. We
observe that for $\infty :1$ the magnetization switches to an angle between
0 and $\pi $ when the current bias is opposite. The origin of this state
clearly differs from previously reported precessional states,\cite%
{kiselev,li2} which required that the applied field is not parallel to the
polarizing (fixed) magnetization. It is a direct consequence of the sign
change in the torque as function of the angle.\cite{manschot} The
switching curves for small deviations from 1 of the different
configurations can be approximated by
$(1-m_z(t))/(1-m_z(0))=\exp(2\alpha\gamma\vert\vect{B}_{\mathrm{eff}}\vert(c-1)t)$.
A smaller damping parameter thus increases 
the switching time but decreases the critical current. The angular dependence of the spin torque
affects the whole switching curve; the torque resembles a sine function for
$1:\infty $, whereas the symmetrical case is closer to Slonczewski's
expression.\cite{slonczewski}

Based on analytic expressions for the spin torque and spin pumping near $%
\theta =0$ in magnetic multilayers we conclude that the critical current for
magnetization reversal in nano-scale spin valves can be  reduced by up to an
order of magnitude by engineering the resistance distribution in the
magnetically active region. The spin torque changes sign for specific
asymmetries giving rise to a new precessional state. After submission
of this article, Jiang \textit{et al.} \cite{jiang} reported a
strongly reduced switching current by modifying the resistance
distribution in the nanopillar by a Ru insertion.  

We would like to thank A.D. Kent for fruitful discussions. This work was
financially supported by FOM, the Norwegian Research Council, and the
NEDO joint research program \textquotedblleft Nano-Scale
Magneto-electronics\textquotedblright.

\newpage 
\setstretch{2}

\bibliographystyle{apsrev}

\begin{thebibliography}{24}
\expandafter\ifx\csname natexlab\endcsname\relax\def\natexlab#1{#1}\fi
\expandafter\ifx\csname bibnamefont\endcsname\relax
  \def\bibnamefont#1{#1}\fi
\expandafter\ifx\csname bibfnamefont\endcsname\relax
  \def\bibfnamefont#1{#1}\fi
\expandafter\ifx\csname citenamefont\endcsname\relax
  \def\citenamefont#1{#1}\fi
\expandafter\ifx\csname url\endcsname\relax
  \def\url#1{\texttt{#1}}\fi
\expandafter\ifx\csname urlprefix\endcsname\relax\def\urlprefix{URL }\fi
\providecommand{\bibinfo}[2]{#2}
\providecommand{\eprint}[2][]{\url{#2}}

\bibitem[{\citenamefont{Slonczewski}(1996)}]{slonczewski}
\bibinfo{author}{\bibfnamefont{J.~C.} \bibnamefont{Slonczewski}},
  \bibinfo{journal}{J. Magn. Magn. Mater.} \textbf{\bibinfo{volume}{159}},
  \bibinfo{pages}{L1} (\bibinfo{year}{1996}).

\bibitem[{\citenamefont{Berger}(1996)}]{berger}
\bibinfo{author}{\bibfnamefont{L.}~\bibnamefont{Berger}},
  \bibinfo{journal}{Phys. Rev. B} \textbf{\bibinfo{volume}{54}},
  \bibinfo{pages}{9353} (\bibinfo{year}{1996}).

\bibitem[{\citenamefont{Oezyilmaz et~al.}(2003)\citenamefont{Oezyilmaz, Kent,
  Monsma, Sun, Rooks, and Koch}}]{kent}
\bibinfo{author}{\bibfnamefont{B.}~\bibnamefont{Oezyilmaz}},
  \bibinfo{author}{\bibfnamefont{A.~D.} \bibnamefont{Kent}},
  \bibinfo{author}{\bibfnamefont{D.}~\bibnamefont{Monsma}},
  \bibinfo{author}{\bibfnamefont{J.~Z.} \bibnamefont{Sun}},
  \bibinfo{author}{\bibfnamefont{M.~J.} \bibnamefont{Rooks}}, \bibnamefont{and}
  \bibinfo{author}{\bibfnamefont{R.~H.} \bibnamefont{Koch}},
  \bibinfo{journal}{Phys. Rev. Lett.} \textbf{\bibinfo{volume}{91}},
  \bibinfo{pages}{67203} (\bibinfo{year}{2003}).

\bibitem[{\citenamefont{Kiselev et~al.}(2003)\citenamefont{Kiselev, Sankey,
  Krivorotov, Emley, Schoelkopf, Buhrman, and Ralph}}]{kiselev}
\bibinfo{author}{\bibfnamefont{S.~I.} \bibnamefont{Kiselev}},
  \bibinfo{author}{\bibfnamefont{J.~C.} \bibnamefont{Sankey}},
  \bibinfo{author}{\bibfnamefont{I.~N.} \bibnamefont{Krivorotov}},
  \bibinfo{author}{\bibfnamefont{N.~C.} \bibnamefont{Emley}},
  \bibinfo{author}{\bibfnamefont{R.~J.} \bibnamefont{Schoelkopf}},
  \bibinfo{author}{\bibfnamefont{R.~A.} \bibnamefont{Buhrman}},
  \bibnamefont{and} \bibinfo{author}{\bibfnamefont{D.~C.} \bibnamefont{Ralph}},
  \bibinfo{journal}{Nature} \textbf{\bibinfo{volume}{425}},
  \bibinfo{pages}{380} (\bibinfo{year}{2003}).

\bibitem[{\citenamefont{Rippard et~al.}(2004)\citenamefont{Rippard, Pufall,
  Kaka, Russek, and Silva}}]{rippard}
\bibinfo{author}{\bibfnamefont{W.~H.} \bibnamefont{Rippard}},
  \bibinfo{author}{\bibfnamefont{M.~R.} \bibnamefont{Pufall}},
  \bibinfo{author}{\bibfnamefont{S.}~\bibnamefont{Kaka}},
  \bibinfo{author}{\bibfnamefont{S.~E.} \bibnamefont{Russek}},
  \bibnamefont{and} \bibinfo{author}{\bibfnamefont{T.~J.} \bibnamefont{Silva}},
  \bibinfo{journal}{Phys. Rev. Lett.} \textbf{\bibinfo{volume}{92}}
  (\bibinfo{year}{2004}).

\bibitem[{\citenamefont{Sun}(2003)}]{sun4}
\bibinfo{author}{\bibfnamefont{J.~Z.} \bibnamefont{Sun}},
  \bibinfo{journal}{Nature} \textbf{\bibinfo{volume}{425}},
  \bibinfo{pages}{359} (\bibinfo{year}{2003}).

\bibitem[{\citenamefont{Brataas et~al.}(2000)\citenamefont{Brataas, Nazarov,
  and Bauer}}]{brataas2}
\bibinfo{author}{\bibfnamefont{A.}~\bibnamefont{Brataas}},
  \bibinfo{author}{\bibfnamefont{Y.~V.} \bibnamefont{Nazarov}},
  \bibnamefont{and} \bibinfo{author}{\bibfnamefont{G.~E.~W.}
  \bibnamefont{Bauer}}, \bibinfo{journal}{Phys. Rev. Lett.}
  \textbf{\bibinfo{volume}{84}}, \bibinfo{pages}{2481} (\bibinfo{year}{2000}).

\bibitem[{\citenamefont{Tserkovnyak et~al.}(2002)\citenamefont{Tserkovnyak,
  Brataas, and Bauer}}]{tserkovnyak2}
\bibinfo{author}{\bibfnamefont{Y.}~\bibnamefont{Tserkovnyak}},
  \bibinfo{author}{\bibfnamefont{A.}~\bibnamefont{Brataas}}, \bibnamefont{and}
  \bibinfo{author}{\bibfnamefont{G.~E.~W.} \bibnamefont{Bauer}},
  \bibinfo{journal}{Phys. Rev. Lett.} \textbf{\bibinfo{volume}{88}},
  \bibinfo{pages}{117601} (\bibinfo{year}{2002}).

\bibitem[{\citenamefont{Xia et~al.}(2002)\citenamefont{Xia, Kelly, Bauer,
  Brataas, and Turek}}]{xia}
\bibinfo{author}{\bibfnamefont{K.}~\bibnamefont{Xia}},
  \bibinfo{author}{\bibfnamefont{P.~J.} \bibnamefont{Kelly}},
  \bibinfo{author}{\bibfnamefont{G.~E.~W.} \bibnamefont{Bauer}},
  \bibinfo{author}{\bibfnamefont{A.}~\bibnamefont{Brataas}}, \bibnamefont{and}
  \bibinfo{author}{\bibfnamefont{I.}~\bibnamefont{Turek}},
  \bibinfo{journal}{Phys. Rev. B} \textbf{\bibinfo{volume}{65}},
  \bibinfo{pages}{220401} (\bibinfo{year}{2002}).

\bibitem[{\citenamefont{Bauer et~al.}(2003)\citenamefont{Bauer, Tserkovnyak,
  Huertas-Hernando, and Brataas}}]{bauer}
\bibinfo{author}{\bibfnamefont{G.~E.~W.} \bibnamefont{Bauer}},
  \bibinfo{author}{\bibfnamefont{Y.}~\bibnamefont{Tserkovnyak}},
  \bibinfo{author}{\bibfnamefont{D.}~\bibnamefont{Huertas-Hernando}},
  \bibnamefont{and} \bibinfo{author}{\bibfnamefont{A.}~\bibnamefont{Brataas}},
  \bibinfo{journal}{Phys. Rev. B} \textbf{\bibinfo{volume}{67}},
  \bibinfo{pages}{94421} (\bibinfo{year}{2003}).

\bibitem[{\citenamefont{Stiles and Zangwill}(2002)}]{stiles}
\bibinfo{author}{\bibfnamefont{M.~D.} \bibnamefont{Stiles}} \bibnamefont{and}
  \bibinfo{author}{\bibfnamefont{A.}~\bibnamefont{Zangwill}},
  \bibinfo{journal}{Phys. Rev. B} \textbf{\bibinfo{volume}{66}},
  \bibinfo{pages}{14407} (\bibinfo{year}{2002}).

\bibitem[{\citenamefont{Manschot et~al.}(2004)\citenamefont{Manschot, Brataas,
  and Bauer}}]{manschot}
\bibinfo{author}{\bibfnamefont{J.}~\bibnamefont{Manschot}},
  \bibinfo{author}{\bibfnamefont{A.}~\bibnamefont{Brataas}}, \bibnamefont{and}
  \bibinfo{author}{\bibfnamefont{G.~E.~W.} \bibnamefont{Bauer}},
  \bibinfo{journal}{Phys. Rev. B} \textbf{\bibinfo{volume}{69}},
  \bibinfo{pages}{092407} (\bibinfo{year}{2004}).

\bibitem[{\citenamefont{Kovalev et~al.}(2002)\citenamefont{Kovalev, Brataas,
  and Bauer}}]{kovalev}
\bibinfo{author}{\bibfnamefont{A.~A.} \bibnamefont{Kovalev}},
  \bibinfo{author}{\bibfnamefont{A.}~\bibnamefont{Brataas}}, \bibnamefont{and}
  \bibinfo{author}{\bibfnamefont{G.~E.~W.} \bibnamefont{Bauer}},
  \bibinfo{journal}{Phys. Rev. B} \textbf{\bibinfo{volume}{66}},
  \bibinfo{pages}{224424} (\bibinfo{year}{2002}).

\bibitem[{\citenamefont{{Pratt Jr.} et~al.}(1991)\citenamefont{{Pratt Jr.},
  Lee, Slaughter, Loloee, Schroeder, and Bass}}]{pratt}
\bibinfo{author}{\bibfnamefont{W.~P.} \bibnamefont{{Pratt Jr.}}},
  \bibinfo{author}{\bibfnamefont{S.~F.} \bibnamefont{Lee}},
  \bibinfo{author}{\bibfnamefont{J.~M.} \bibnamefont{Slaughter}},
  \bibinfo{author}{\bibfnamefont{R.}~\bibnamefont{Loloee}},
  \bibinfo{author}{\bibfnamefont{P.~A.} \bibnamefont{Schroeder}},
  \bibnamefont{and} \bibinfo{author}{\bibfnamefont{J.}~\bibnamefont{Bass}},
  \bibinfo{journal}{Phys. Rev. Lett.} \textbf{\bibinfo{volume}{66}},
  \bibinfo{pages}{3060} (\bibinfo{year}{1991}).

\bibitem[{\citenamefont{Gijs et~al.}(1993)\citenamefont{Gijs, Lenczowski, and
  Giesbers}}]{gijs}
\bibinfo{author}{\bibfnamefont{M.~A.~M.} \bibnamefont{Gijs}},
  \bibinfo{author}{\bibfnamefont{S.~K.~J.} \bibnamefont{Lenczowski}},
  \bibnamefont{and} \bibinfo{author}{\bibfnamefont{J.~B.}
  \bibnamefont{Giesbers}}, \bibinfo{journal}{Phys. Rev. Lett.}
  \textbf{\bibinfo{volume}{70}}, \bibinfo{pages}{3343} (\bibinfo{year}{1993}).

\bibitem[{\citenamefont{Yang et~al.}(1995)\citenamefont{Yang, Holody, Loloee,
  Henry, {Pratt Jr.}, Schroeder, and Bass}}]{yang}
\bibinfo{author}{\bibfnamefont{Q.}~\bibnamefont{Yang}},
  \bibinfo{author}{\bibfnamefont{P.}~\bibnamefont{Holody}},
  \bibinfo{author}{\bibfnamefont{R.}~\bibnamefont{Loloee}},
  \bibinfo{author}{\bibfnamefont{L.~L.} \bibnamefont{Henry}},
  \bibinfo{author}{\bibfnamefont{W.~P.} \bibnamefont{{Pratt Jr.}}},
  \bibinfo{author}{\bibfnamefont{P.~A.} \bibnamefont{Schroeder}},
  \bibnamefont{and} \bibinfo{author}{\bibfnamefont{J.}~\bibnamefont{Bass}},
  \bibinfo{journal}{Phys. Rev. B} \textbf{\bibinfo{volume}{51}},
  \bibinfo{pages}{3226} (\bibinfo{year}{1995}).

\bibitem[{\citenamefont{Lifshitz and Pitaevskii}(1980)}]{landau}
\bibinfo{author}{\bibfnamefont{E.~M.} \bibnamefont{Lifshitz}} \bibnamefont{and}
  \bibinfo{author}{\bibfnamefont{L.~P.} \bibnamefont{Pitaevskii}},
  \emph{\bibinfo{title}{Statistical Physics, Part 2}}
  (\bibinfo{publisher}{Pergamon Press}, \bibinfo{year}{1980}).

\bibitem[{\citenamefont{Gilbert}(1955)}]{gilbert}
\bibinfo{author}{\bibfnamefont{T.~L.} \bibnamefont{Gilbert}},
  \bibinfo{journal}{Phys. Rev.} \textbf{\bibinfo{volume}{100}},
  \bibinfo{pages}{1243} (\bibinfo{year}{1955}).

\bibitem[{\citenamefont{Sun}(2000)}]{sun2}
\bibinfo{author}{\bibfnamefont{J.~Z.} \bibnamefont{Sun}},
  \bibinfo{journal}{Phys. Rev. B} \textbf{\bibinfo{volume}{62}}
  (\bibinfo{year}{2000}).

\bibitem[{\citenamefont{Li and Zhang}(2004)}]{li}
\bibinfo{author}{\bibfnamefont{Z.}~\bibnamefont{Li}} \bibnamefont{and}
  \bibinfo{author}{\bibfnamefont{S.}~\bibnamefont{Zhang}},
  \bibinfo{journal}{Phys. Rev. B} \textbf{\bibinfo{volume}{69}},
  \bibinfo{pages}{134416} (\bibinfo{year}{2004}).

\bibitem[{\citenamefont{Tserkovnyak et~al.}(2003)\citenamefont{Tserkovnyak,
  Brataas, and Bauer}}]{tserkovnyak3}
\bibinfo{author}{\bibfnamefont{Y.}~\bibnamefont{Tserkovnyak}},
  \bibinfo{author}{\bibfnamefont{A.}~\bibnamefont{Brataas}}, \bibnamefont{and}
  \bibinfo{author}{\bibfnamefont{G.~E.~W.} \bibnamefont{Bauer}},
  \bibinfo{journal}{Phys. Rev. B} \textbf{\bibinfo{volume}{67}},
  \bibinfo{pages}{140404} (\bibinfo{year}{2003}).

\bibitem[{\citenamefont{Valet}(2004)}]{valet}
\bibinfo{author}{\bibfnamefont{T.}~\bibnamefont{Valet}} (\bibinfo{year}{2004}),
  \bibinfo{note}{unpublished}.

\bibitem[{\citenamefont{Li and Zhang}(2003)}]{li2}
\bibinfo{author}{\bibfnamefont{Z.}~\bibnamefont{Li}} \bibnamefont{and}
  \bibinfo{author}{\bibfnamefont{S.}~\bibnamefont{Zhang}},
  \bibinfo{journal}{Phys. Rev. B} \textbf{\bibinfo{volume}{68}},
  \bibinfo{pages}{024404} (\bibinfo{year}{2003}).

\bibitem[{\citenamefont{Jiang et~al.}(2004)\citenamefont{Jiang, Abe, Ochiai,
  Nozaki, Hirohata, Tezuka, and Inomata}}]{jiang}
\bibinfo{author}{\bibfnamefont{Y.}~\bibnamefont{Jiang}},
  \bibinfo{author}{\bibfnamefont{S.}~\bibnamefont{Abe}},
  \bibinfo{author}{\bibfnamefont{T.}~\bibnamefont{Ochiai}},
  \bibinfo{author}{\bibfnamefont{T.}~\bibnamefont{Nozaki}},
  \bibinfo{author}{\bibfnamefont{A.}~\bibnamefont{Hirohata}},
  \bibinfo{author}{\bibfnamefont{N.}~\bibnamefont{Tezuka}}, \bibnamefont{and}
  \bibinfo{author}{\bibfnamefont{K.}~\bibnamefont{Inomata}},
  \bibinfo{journal}{Phys. Rev. Lett.} \textbf{\bibinfo{volume}{92}},
  \bibinfo{pages}{167204} (\bibinfo{year}{2004}).
\end{thebibliography}

\pagestyle{empty}
\newpage
{\Large FIGURE CAPTIONS}

\begin{itemize}
\item [Figure 1] $m_{z}$ as function of time for switching in magnetic multilayers
with different resistance distributions. The legend denotes the ratio of $%
G_{1}:G_{2}$.
\end{itemize}

\newpage
{\Large TABLE CAPTIONS}
\begin{itemize}
\item [TABLE I.] The slope of the spin torque at $\protect\theta \approx 0$, the
increase of the damping by spin pumping and the critical current for
different asymmetrical configurations of the investigated finite element
system.
\end{itemize}

\newpage

\newpage

\begin{table}[th]
\caption{}
\label{tab:k}
\begin{center}
\begin{tabular}{@{\extracolsep{0.5cm}}|l|r r r|}
\hline
$G_1:G_2$ & $k$ & $\Delta \alpha$ & $I_\mathrm{c,c}$ \\ \hline
$1:\infty$ & 0.565 & 0.0054 & $0.55\,\mathrm{mA}$ \\ 
$1:1$ & 0.117 & 0.0062 & $2.86\,\mathrm{mA}$ \\ 
$\infty:1$ & -0.331 & 0.0131 & $-1.59\,\mathrm{mA}$ \\ \hline
\end{tabular}
\end{center}
\end{table}

\end{document}